# High-resolution single-shot ultrafast imaging at ten trillion frames per second

Xuanke Zeng[1], Jingzhen Li[1]*, Yi Cai[1], Shuiqin Zheng[1], Hu Long[1], Xiaowei Lu[1], Xiuwen Zhang[1], Weixin Xie[2], and Shixiang Xu[1]*

**Ultrafast imaging is a powerful tool for studying space-time dynamics in photonic material, plasma physics, living cells, and neural activity. Pushing the imaging speed to the quantum limit could reveal extraordinary scenes about the questionable quantization of life and intelligence, or the wave-particle duality of light. However, previous designs of ultrafast photography are intrinsically limited by framing speed. Here, we introduce a new technique based on a multiple non-collinear optical parametric amplifier principle (MOPA), which readily push the frame rate into the area of ten trillion frames per second with higher spatial resolution than 30 line pairs per millimeter. The MOPA imaging is applied to record the femtosecond early evolution of laser-induced plasma grating in air for the first time. Our approach avoids the intrinsic limitations of previous methods, thus can be potentially optimized for higher speed and resolution, opening the way of approaching quantum limits to test the fundamentals.**

To get the space-time information of the ultrafast process accurately and reveal its dynamic law by using of ultrahigh speed imaging[1-6], the capabilities of temporal and spatial resolutions of imaging system are two important indicators[7-9]. Therefore, it is the focus of research in this field to realize imaging systems with high temporal resolution, spatial resolution and real-time function[10-12]. Ultrafast processes can be divided into two categories, one of which is a kind of reproducible events, and the other one is a kind of non-repetitive or random events. While the pump-probe method[13-15] is usually adopted for the former one. For the latter one, a single-shot measurement is required to acquire the sequentially timed image information of the events. In recent years, the rapid development of ultrashort pulse laser and the exploration of new principles and techniques have greatly promoted the development of ultrahigh speed imaging and circumvented the limitation of traditional imaging in a single-shot. Serial time-encoded amplified imaging/microscopy (STEAM) is an ultrafast continuous imaging method[16] that runs at a frame rate of ~$10^8$ frames per second (fps) and opens the window onto nanosecond two-dimensional imaging with a shutter speed of ~100 picoseconds (ps). Based on computational imaging, compressed ultrafast photography (CUP) can get a photography frequency of $10^{11}$ fps in a single shot[7]. It breaks the digitalization bandwidth limitation by incorporating a compressed-sensing-based algorithm into the data acquisition of a streak camera with image tube. But the spatial resolution is lower than 1 line pairs per millimeter (lp/mm). By use of tomographic imaging methods, the single-shot multispectral tomography (SMT) and the frequency-domain tomography (FDT) can acquire tomographic images with the temporal resolutions of several picoseconds (corresponding to the rate of the order of $10^{11}$ fps) in a single shot[8,9]. However, the spatial resolutions of SMT and FDT are limited by the divergence angle of each fan beam and limited to imaging of simple structured objects because of the limited number of illumination angles, respectively. Sequentially timed all-optical mapping photography (STAMP)[17] is an ultrafast burst imaging method with a high pixel resolution (450×450 pixels) and a record high frame rate of ~$4.4 \times 10^{12}$ fps, which is the reported highest rate in a single-shot at present. However, due to the use of spectral coding for framing, the framing time and the exposure time and the number of frames are mutually affected. So when the framing time is up to 229 femtoseconds (fs), the exposure time is ~800fs, which results in the modified temporal qualify factor[18,19] $g^{2/3}$ is less than 1. It means that the information between the adjacent frames is severely overlapping and the temporal resolution is insufficiency at such high frame rate. By utilizing spectral filtering in STAMP[20,21], a flexible increase in the number of frame is realized. But the frame rate dropped to $2.1 \times 10^{12}$ fps in order to avoid the overlap of information in STAMP.

In this paper we propose and demonstrate a single-shot burst imaging technique by utilizing MOPA. The key feature of MOPA is that the framing time, only dependent on the relative delays of the pump beams and independent of the exposure time and number of frame, can reach tens of femtosecond. The other feature is that it can obtain a high spatial resolution by optimizing the spatial frequency bandwidth of OPA. Consequently, it pushes the frame rate into the area of ten trillion frames per second ($10^{13}$ fps) with higher spatial resolution than 30 lp/mm in our experiment.

[1] Institute of Photonic Engineering, College of Electron Sci. & Tech., Shenzhen University, Shenzhen Key Laboratory of Micro Nano Photonic Information, Shenzhen, 518060, China. [2] College of Information Engineering, Shenzhen University, Shenzhen, 518060, China. * e-mail: lijz@szu.edu.cn, and shxxu@szu.edu.cn.

# Experimental setup

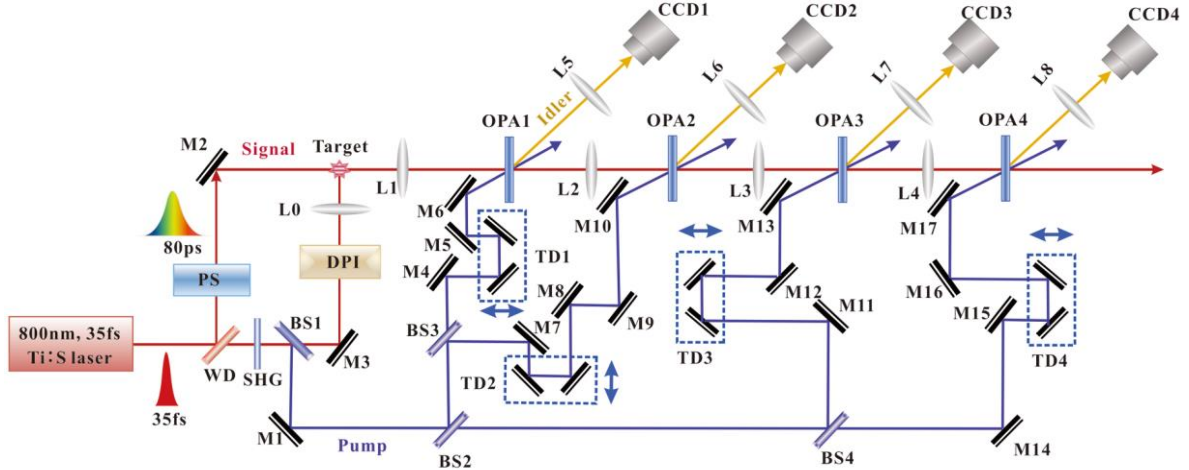

**Figure 1 | Setup for ultrafast real-time MOPA imaging.** WD: Wedge plate; PS: Pulse stretcher; SHG: Second harmonic generator; BS1~BS4: Wavelength separators; DPI: Double pulse interferometer; OPA1~OPA4: optical parametric amplifiers; M1~M17: Mirrors; MG1~MG4: temporal delay lines; L1~L8: Optical imaging lenses; CCD1~CCD4: Charge-coupled devices.

Figure 1 shows our experimental setup for the MOPA imaging to get four sequentially images in a single-shot. The light source is a commercial Ti: sapphire laser system with a pulse width of 35fs and energy of 3.3mJ. 1% energy is separated by a wedge plate and then stretched by a pulse stretcher (PS) as a signal beam. The rest 99% energy is used to produce its second harmonic beam as the pump beams for the four OPAs after a second harmonic generator (SHG) with the conversion efficiency of 30% and three separators (BS2~BS4) with the ratio of 1:1. The remaining 800nm fundamental beam is separated from the second harmonic beam by a wavelength splitter (BS1), and then focus on the object plane of the OPA1 imaging system to produce an air plasma grating as a target after a double pulse interferometer (DPI). The 800nm signal beam, which is a chirp pulse with the pulse width of ~80ps after the PS, illuminates on the plasma grating. In order to meet the imaging needs of the target objects with spatial feature of tens of microns, the signal beam image has a magnified of two times by the first imaging lens L1 (a confocal imaging system with two lenses) from the object plane to the image plane. And then reaches together with the pump beam inside the OPA1 to generate an idler beam separating spatially from both of them with a small intersection angle (~2°). The idler wavelength-converted image is recorded by a CCD camera (CCD1) with a pixel resolution of 1628× 1236 pixels behind a lens (L3) which is used to image the optical information from the rear surface of the OPA1 to the CCD1 with two-fold reduction. As the idler is generated only during the 35fs of the amplification process depending on the pump pulse width, the chirp signal beam reaching the crystal before or after the pump pulse is not amplified. Hence, a very good shutter is obtained. Similar to the OPA1 imaging, we get three other idler images by three other OPAs (OPA2~OPA4, the same as OPA1). Each OPA is placed on the image plane successively. As a result, we can get four idler images from the four OPAs by four different CCD cameras (CCD1~CCD4) respectively in a single-shot.

In this setup, the framing time, which depends on the relative delays of the pump beams, can reach tens of fs just by adjusting each pump delay (TD1~TD4) of the four pump beams after a calibration of 0 moment of the four OPAs because each image comes from an independent idler beam of each OPA. The exposure time, which is measured by a SPIDER setup as shown in Fig.2a, is approximately equal to 40fs because of a good shutter is obtained by the idler beam after OPA. As the framing time $\Delta t$=100fs (corresponding to the frame rate of $10^{13}$fps) is obtained in the following experimental data, the modification temporal factor $g^{2/3}$ can be calculated to be 1.84 which means no-overlap of information between adjacent frames. And the frame rate can be further improved by optimizing the MOPAs imaging system. In addition, the number of frames can be increased freely without affecting the framing time and exposure time by designing a more compact optical system theoretically.

**High spatial resolution of ultrafast MOPA imaging**
The spatial resolution is one of the important indicators of imaging, but it is often sacrificed in order to obtain high temporal resolution in the previous ultrafast imaging, e.g., CUP, and FDT. The MOPA circumvents the limitation of spatial resolution at the frame rate as high as $10^{13}$fps. As we know, the spatial frequency bandwidth[22-26], which depends on the crystal thickness and pump energy, is the main factor affecting the spatial resolution in OPA imaging. Fig.2b shows the parametric gain versus spatial frequency of a 2-mm-thick type-I phase-matching $\beta$-BBO at a pump level of 30GW/cm$^2$ (the solid line) and a 0.5-mm-thick type-I phase-matching $\beta$-BBO at a pump level of 10GW/cm$^2$ (the dotted line). In our previous work[27], a type-I *β*-BBO crystal with the length of 2mm is used to image a static target object of a USAF 1951(0) test pattern. The gain bandwidth at 3dB is estimated to 28 lp/mm from the solid line of Fig.2B. We recorded an idler image which has the largest visible spatial feature of 22.62 lp/mm in the vertical direction with an imaging area of 6.8 × 6.8 mm$^2$. In order to further optimize the spatial resolution, a thinner *β*-BBO crystal with the length of 0.5mm is used for OPA imaging in the ultrafast MOPA imaging setup, the spatial gain bandwidth at 3dB of which is promoted to 40 lp/mm (the dotted line in Fig.2b) much higher than that of 28 lp/mm by using the thicker crystal with the 2mm crystal. Consequently, we get an idler image has the largest visible spatial feature of 36 lp/mm (element 5.2) in the vertical direction with a high pixel resolution of 1200×1000 pixels as shown in Fig.2c. Here, the idler beam image has a ratio of 1:1 from the object plane to the plane of CCD camera. Since each image is recorded on a separate camera for the multiple imaging, it is easy to obtain a large space-bandwidth-product in the MOPA imaging system.

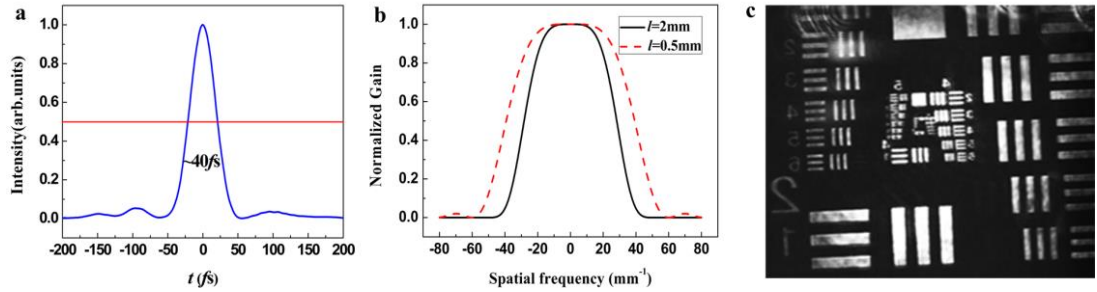

**Figure 2 | a**, The measurement of the exposure time Δτ by a SPIDER setup. **b**, The parametric gains vs. spatial frequency of a non-collinear type-I phase-matching *β*-BBO OPA. **c**, The idler image of the target USAF 1951(0) by using a 0.5-mm-thick type-I phase-matching *β*-BBO.

**Experimental movies of the Ultrafast real-time MOPAs imaging**

The measured target is a laser-induced plasma grating in air which is formed by two equal beams, synchronizing in time and focusing with a crossing angle on the object plane of the first OPA imaging system (in Fig.1). When the crossing angle is set to 2.6°, the laser-induced plasma grating has a spatial structure of five stripes with the stripe's spacing of 17μm between peak to peak in the horizontal direction. The movie frames of the plasma grating at four different moments, which is obtained by the ultrafast MOPA imaging in a single-shot, is shown in Fig.3. We firstly identified a 0 moment (shown in Fig. 3a of OPA1) for calibration of the four idler images, and then regulated the relative pump delays of the second, third and fourth OPA to 100, 200, and 300fs after the 0 moment in turn by precision displacement platforms. As a result, four instantaneous idler images of the plasma grating early evolution at the frame rate of 1×10$^{13}$fps were recorded by four CCD cameras with a pixel resolution of 600×600 pixels. One can see that the stripes of the plasma gradually appeared from the left to the right in images, and their definition to have a rise. This shows that within the moment of 0~300fs, the free electron density of the plasma grating is increasing monotonously. Since the frame intervals depend only on four independent pump delays and can be adjusted freely, we can obtain a motion-picture not only with equally frame intervals, but also with unequally frame intervals according to demand. Fig.3b and Fig.3c show movie frames of the plasma grating with an unequally frame interval for the moment from 0.8 to 4ps (0.8, 1, 2, and 4ps), and from 6 to 30ps (6, 10, 20, and 30ps), respectively. The stripes definition of plasma grating in Fig.3B gradually increases from OPA1 to OPA4, and is better than in Fig.3a. So we can get that the plasma density gradually increased after the moment of 300fs. That is to say, the electron density of the plasma grating is still increasing from the time of 0.8 to 4ps. In the later evolution of plasma grating, the stripes definition of the plasma appears to have a decline trend from 6 to 30ps as shown in Fig.3c. We can get that the refractive index, or the electron density of plasma attenuates gradually with the pump delay time in this process. It shows that the lifetime of the excited plasma grating is about 30ps because the stripes are difficult to distinguish at the moment of 30ps.

To further illustrate the plasma grating image changes with the relative pump delay and the ability of spatial resolution, we calculate the plasma grating stripe modulation index, which is defined as

$$M = \frac{I_{\max} - I_{\min}}{I_{\max} + I_{\min}} \quad (1)$$

where $I_{max}$ and $I_{min}$ are the maximum and minimum intensity of the stripe. Owing to the profiles of the four pump beams are not exactly the same in experiment, the frames of OPAs in Fig.3 are different. In order to eliminate the influence of back noise from different idler images on the calculation of the modulation index, we measured the pump intensity distribution of the four OPAs respectively, and blocked the plasma to get idler background images of OPAs without the object input, and then use the difference between the images with and without the object input signal to reduce the influence of the different profiles of pump beams. The normalized gray values in the direction perpendicular to the stripes of the four images in Fig.3a is shown in Fig.4a. It shows the idler intensity variation from pixel to pixel for the four moments by solid lines with different colors. The five strips are visible with the stripe's spacing of ~17μm between peak to peak in the horizontal direction, and the modulations change with time. Combining multiple measurements, the curve of the third strip's modulation indexes change with time is calculated and showed in Fig. 4b. The modulation index rapidly reaches a comparatively high value after the plasma excitation at 0 moment, it increases monotonously and reaches the maximum value rapidly at the moment of 4ps, and then it begins to decline gradually and approaches to 0 at 30ps. This coincides with the observations of early and later evolution processes of plasma grating in air in Fig.3.

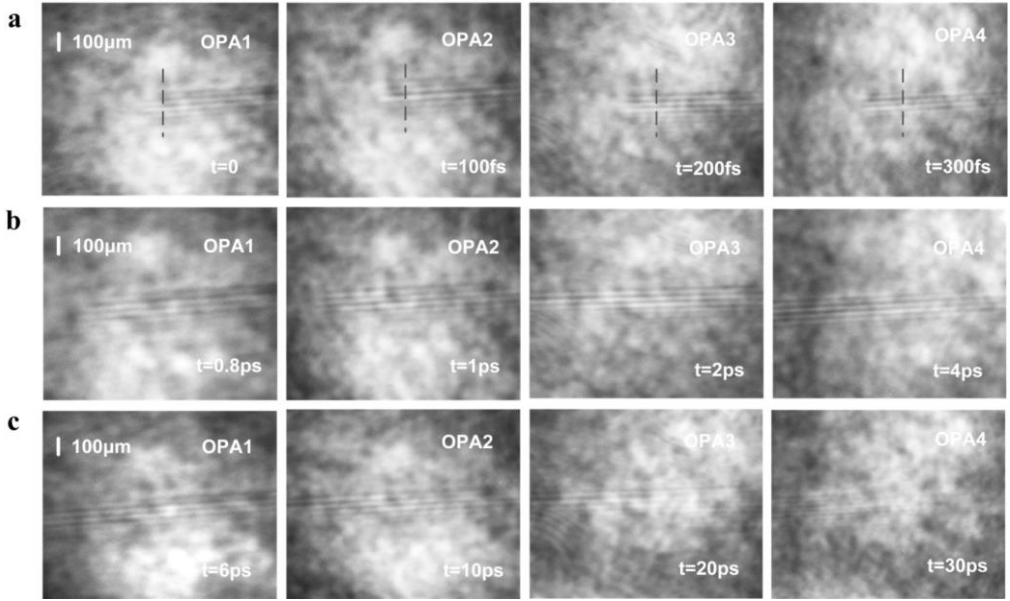

**Figure 3 | a,** Four instantaneous idler images of the plasma grating early evolution at the frame rate of 1×10$^{13}$fps (t=0, 100, 200, and 300fs). **b,** Four frames with an unequally frame interval for the moment of 0.8, 1, 2, and 4ps. **c,** Four frames of the later evolution for the moment of 6, 10, 20, and 30ps.

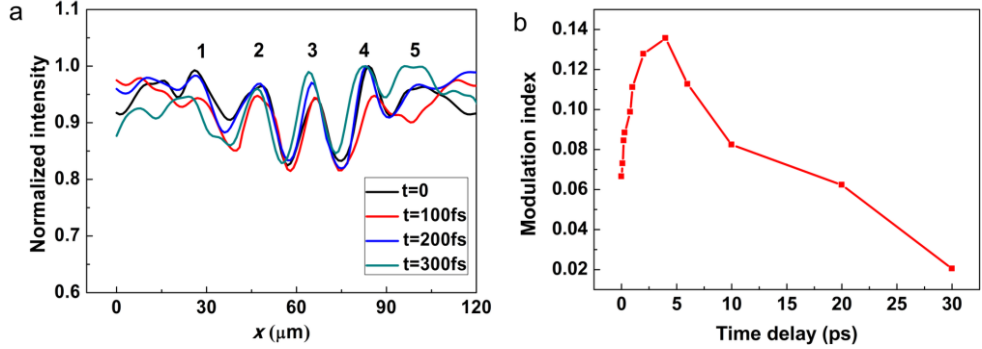

**Figure 4 | a,** The normalized gray values in the direction perpendicular to the stripes for the moments from 0 to 300fs. **b,** The third stripe modulation indexes versus pump time delay of the idler images from 0 to 30ps.

In summary, we have introduced the new technique based on a multiple non-collinear optical parametric amplifier principle and demonstrated a novel single-shot ultrafast imaging for visualizing spatiotemporal dynamics of a plasma grating. The highest frame rate of the motion-pictures is up to 10$^{13}$ fps with the high spatial resolution of more than 30 lp/mm in the experiment. The frame rate, which depends on the accuracy of the displacement platform and the width of pump pulse, can be further improved by optimizing the system according to the needs of imaging. This kind of imaging breaks through the intrinsic limitations of previous methods and provides a new way for blur-free observation of fast transient dynamics including plasma physics, photochemistry and biomedical medicine at the atom level.

## Methods

**Spatial frequency bandwidth of OPA imaging.** The gain for un-depleted plane pump approximation can be written as[23]

$$G_s = |\mu_q|^2, \quad G_i = |\nu_q|^2 \quad (2)$$

where the subscripts '$s$' and '$i$' stand for the signal and idler, $q$ is for spatial frequency. The coupling coefficients $\mu$ and $\nu$ are given by

$$\mu(q) = \left[\cosh(hl) + i\frac{\Delta k_{eff} l}{2}\frac{\sinh(hl)}{hl}\right] \times \exp\left(-i\frac{\Delta k_{eff}}{2}l\right),$$
$$\nu(q) = \left[i\frac{\kappa l}{2}\frac{\sinh(hl)}{hl}\right] \times \exp\left(i\frac{\Delta k_{eff}}{2}l\right), \quad (3)$$

where $h = [\kappa^2-(\Delta k_{eff})2]^{1/2}/2$, $\Delta k_{eff}$ = k$_p$-k$_s$-k$_i$+2π$^2$q$^2$ (k$_s^{-1}$+k$_i^{-1}$) is the effective phase mismatch, $\kappa$ is the parametric gain coefficient which is proportional to the intensity of the pump beam, and l is the length of parametric crystal. The spatial frequency bandwidth $\Delta f$ of OPA gain at 3dB can be written as [24]

$$\Delta f = \frac{1}{\pi}\left(\frac{k_i^2 \kappa \ln 2}{4l}\right)^{\frac{1}{4}}. \quad (4)$$

The equation (2) ~ (4) shows that thin crystal thickness with strong pump intensity is helpful to realize a large spatial frequency bandwidth for OPA imaging. However, excessive pump intensity will damage the crystal, while thin crystal thickness will make the gain reduction.

**Generation of plasma grating.** The measured target is a plasma grating which is formed by focusing two equal femtosecond pulse beams with a small crossing angle. The diagram is shown in extended data Fig.1a. As the crossing angle becomes larger, the number of stripes increases, while its spacing becomes smaller. As shown in extended data Fig. 1b, there are three stripes in the picture and the distance between peak to peak is about 30μm when 2α=1.5°. In order to embody the spatial resolution of the MOPA, we adjust the crossing angle of the double interference pulses from 1.5 ° to 2.6 ° to form a target of finer spatial structure. As shown in extended data Fig. 1c, the measured target, a laser-induced plasma grating, has a spatial structure of five stripes with the distance between peak to peak of 17μm in the horizontal direction. The stripe's spacing of extended data Fig. 1b and 1c can be estimated by the distribution of gray value along the vertical direction of the stripes in extended data Fig. 1d.

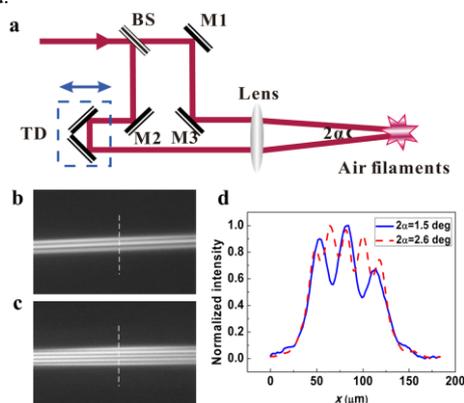

**Extended Data Figure 1 | a**, Diagram of DPI for generating ultrafast events. **b**, The laser-induced plasma grating in air by two femtosecond pulses focusing with a crossing angle of 1.5°. **c**, The laser-induced plasma grating in air by two femtosecond pulses focusing with a crossing angle of 2.6°. **d**, the normalized gray value in the direction perpendicular to the stripes in Fig. 1b and 1c.

## Acknowledgments
The authors thank Prof. Zenghu Chang, a famous scientist in the field of attosecond laser pulse at the University of Central Florida, USA, for many times discussions. This work was supported by National Natural Science Fund of China (61775142, 61705132，61027014), the China Postdoctoral Science Foundation (2017M612726), and the Specialized Research Fund for the Shenzhen Strategic Emerging Industries Development (JCYJ20150324141711651, and JCYJ20150525092941064).

## Author contributions
X.Zeng, J.Li, and S.Xu developed the concept and design of the experiments, Y.Cai, S.Zheng and X.Lu carried out the numerical analysis. X.Zeng., Y.Cai., and S.Zheng. conducted the experiments, X.Zeng, J.Li, S.Xu, H.Long and X.Zhang wrote the main manuscript text, and W.Xie gave help guidance and discussions. All authors discussed the results and reviewed the manuscript.

## Author information
Reprints and permissions information is available at www.nature.com/reprints. The authors declare no competing financial interests. Correspondence and requests for materials should be addressed to J.Li (lijz@szu.edu.cn), and S.Xu (shxxu@szu.edu.cn).